\documentclass[aps,pre,twocolumn,groupedaddress]{revtex4-2}

\usepackage{graphicx}
\usepackage{amsmath}
\usepackage{amsthm,amssymb}
\usepackage{mathtools}

\newtheorem{theorem}{Theorem}

\newtheorem{corollary}{Corollary}
\theoremstyle{definition}
\newtheorem{example}{\bf Example}
\newtheorem{remark}{\bf Remark}

\begin{document}


\title{Explicit kinematic equations for degree-4 rigid origami vertices, \\ Euclidean and non-Euclidean}


\author{Riccardo Foschi}
\affiliation{University of Bologna, DA, Via Risorgimento 2, 40136 Bologna, Italy}
\email{riccardo.foschi2@unibo.it}

\author{Thomas C. Hull}
\affiliation{Western New England University, 1215 Wilbraham Road, Springfield, MA 01119, USA}
\email{thull@wne.edu}

\author{Jason S. Ku}
\affiliation{National University of Singapore, 9 Engineering Drive 1, \#07-08 Block EA, Singapore 117575}
\email{jasonku@nus.edu.sg}


\date{\today}

\begin{abstract}

We derive new algebraic equations for the folding angle relationships in completely general degree-four rigid-foldable origami vertices, including both Euclidean (developable) and non-Euclidean cases. These equations in turn lead to novel, elegant equations for the general developable degree-four case. We compare our equations to previous results in the literature and provide two examples of how the equations can be used: In analyzing a family of square twist pouches with discrete configuration spaces, and for proving that a new folding table design made with hyperbolic vertices has a single folding mode.

\end{abstract}


\maketitle

\section{Introduction}

\begin{figure}
    \centering
    \includegraphics[scale=.25]{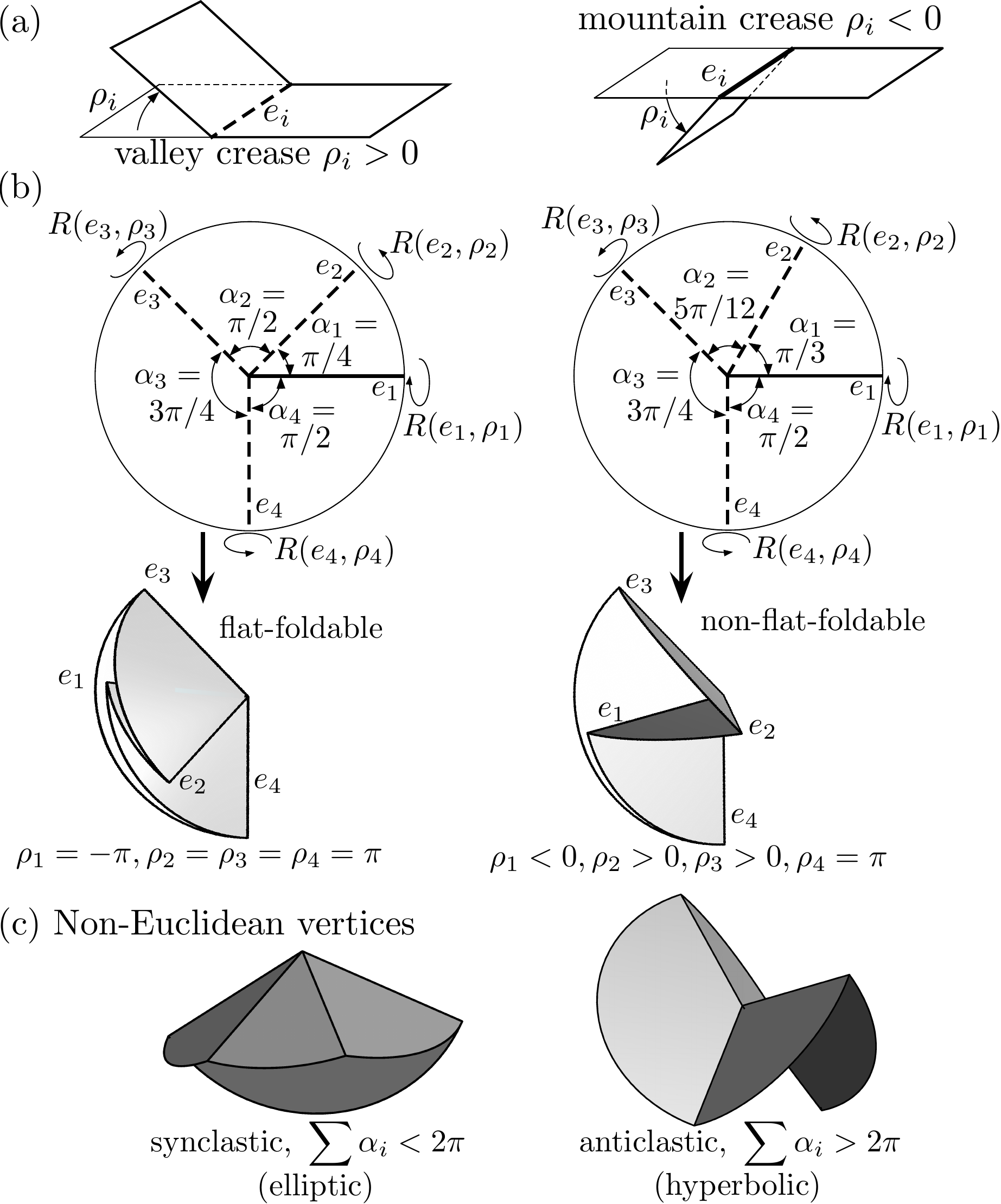}
    \caption{(a) Folding angles $\rho_i$ define valley and mountain creases. (b) Sector angles $\alpha_i$ defining examples of flat- and non-flat-foldable vertices. (c) Examples of non-Euclidean vertices.}
    \label{fig0}
\end{figure}

The folding of stiff, two-dimensional materials along straight crease line segments so that the material remains planar between creases is commonly known as \textit{rigid-foldable origami}. It has garnered the attention of designers, engineers, and physicists as a source of easy-to-manufacture, collapsible mechanisms for use in everything from metamaterials to solar sail deployment in space to furniture design \cite{Fulton21,YOrigami,Silverberg1,YouSci14}.  Of particular interest have been rigid origami structures that flex with only a single degree of freedom, thus giving controllable folding mechanics. One way to study such mechanics is to quantify the \textit{folding angle} at each crease as the origami structure flexes. A folding angle $\rho_i$ is the amount the material deviates from a flat, unfolded state at a crease $e_i$; see Figure~\ref{fig0}(a). When four creases meet at a vertex, as in Figure~\ref{fig0}(b), the folding mechanism will have one degree of freedom, meaning that one crease's folding angle will determine the folding angles of the other three creases. Finding equations for these determined folding angles in terms of the indeterminate folding angle has been an essential part of many studies of rigid origami in applications \cite{Fang16,Feng20,Tachi15,Silverberg2}. Such folding angle equations provide a pure mathematical model of rigid folding that ignores thickness of the material or bending energy at the creases. They nonetheless provide valuable information on the configuration and relative speeds of the folded creases as the mechanism flexes.

In this paper we provide folding angle equations that hold for the full range of possibilities for a degree-4 rigid origami vertex (i.e., where four crease lines meet). The \textit{flat-foldable} case, where the folding angles may flex to the point where they all equal $\pm \pi$ (see Figure~\ref{fig0}(b) left) has very elegant folding angle equations that are well known; these will be summarized along with other background material in Section~\ref{sec2}.  In Section~\ref{sec3} we present our new equations which not only cover the flat- and non-flat-foldable degree-4 vertex cases, but also work for the so-called \textit{non-Euclidean} degree-4 vertex cases where the sum of the sector angles $\alpha_i$ between adjacent creases on the folded material do not sum to $2\pi$ (see Figure~\ref{fig0}(c). These new, fully-general equations lead to novel and surprisingly elegant equations for the Euclidean, non-flat-foldable degree-4 case, such as the one shown in Figure~\ref{fig0}(b) right; this will be the subject of Section~\ref{sec4}. Finally, in Section~\ref{sec5} we use our new equations in two applications: (A) a family of twist-based origami pouches that have finite, disconnected rigid origami configuration spaces and therefore exhibit bistability by ``snapping" into their target form when folded; and (B) the design of a foldable table with hyperbolic vertices.

\section{Background on folding degree-4 vertices}\label{sec2}

We define the \textit{crease pattern} of a rigid-folding origami to be the planar graph of straight line segments drawn on the material that is to be folded.  In this paper we focus on crease patterns that have only one vertex in the material's interior, sometimes called \textit{single-vertex} crease patterns. The angles between consecutive creases at the vertex on the unfolded material are called \textit{sector angles}. We denote the sector angle between creases $e_i$ and $e_{i+1}$ by $\alpha_i$ (where the indices are taken cyclically, mod 4 for a degree-4 vertex).  A single-vertex crease pattern whose sector angles sum to $2\pi$ is called \textit{developable}, aka Euclidean. 
Some of the many studies of the kinematics of degree-4, developable origami vertices are \cite{Fang16,Waitukaitis16,Stern17,Luca19,Feng20,He20,Jigsaw20}.  More recent studies have explored rigid foldings of \textit{non-developable}, aka non-Euclidean degree-4 vertices \cite{Toptrans20,NEO20}, where $\sum\alpha_i\ne 2\pi$.  Non-Euclidean origami vertices come in two types: the \textit{synclastic case} of an elliptic, convex polyhedral cone crease pattern where the sector angles have $\sum\alpha_i < 2\pi$ and the \textit{anticlastic case} of a hyperbolic vertex with $\sum\alpha_i > 2\pi$ (see Figure~\ref{fig0}(c)). 

We denote the folding angle of a crease $e_i$ by $\rho_i$.  If $\rho_i>0$ we call $e_i$ a \textit{valley} crease, whereas if $\rho_i<0$ it is called a \textit{mountain} crease. As seen in Figure~\ref{fig0}(a) and (b), valley creases are denoted in illustrations by a dashed line, while mountains are drawn with a solid bold line. 

If we can fold an origami crease pattern to a point where it lies in a plane, with all the folding angles equal to $\pi$ or $-\pi$, then we say that the crease pattern is \textit{flat-foldable}. One of the basic results of flat-foldable origami is \textit{Kawasaki's Theorem}, which states that a necessary and sufficient condition for a degree-4 vertex to be flat-foldable is that the sector angles between creases satisfy $\alpha_1-\alpha_2+\alpha_3-\alpha_4=0$ \cite{Hull1,Origametry}.  

We define the \textit{configuration space} of a degree-4 rigid-foldable origami vertex $V$ to be the set of points 
$(\rho_1, \rho_2, \rho_3, \rho_4)\in\mathbb{R}^4$ such that $V$ can be rigidly folded with folding angles $\rho_i$ at each crease $e_i$. If we let $R(e_i, \rho_i)$ denote the orthogonal matrix that rotates $\mathbb{R}^3$ about the line containing crease $e_i$ by angle $\rho_i$, then a necessary condition for an origami vertex to be rigidly foldable with folding angles $\rho_i$ is $\prod R(e_i,\rho_i)=I$ where $I$ is the identity matrix \cite{belcastro}.  The action of the matrices $R(e_i,\rho_i)$ is shown in the degree-4 crease patterns of Figure~\ref{fig0}(b).

\begin{figure*}
\includegraphics[width=\linewidth]{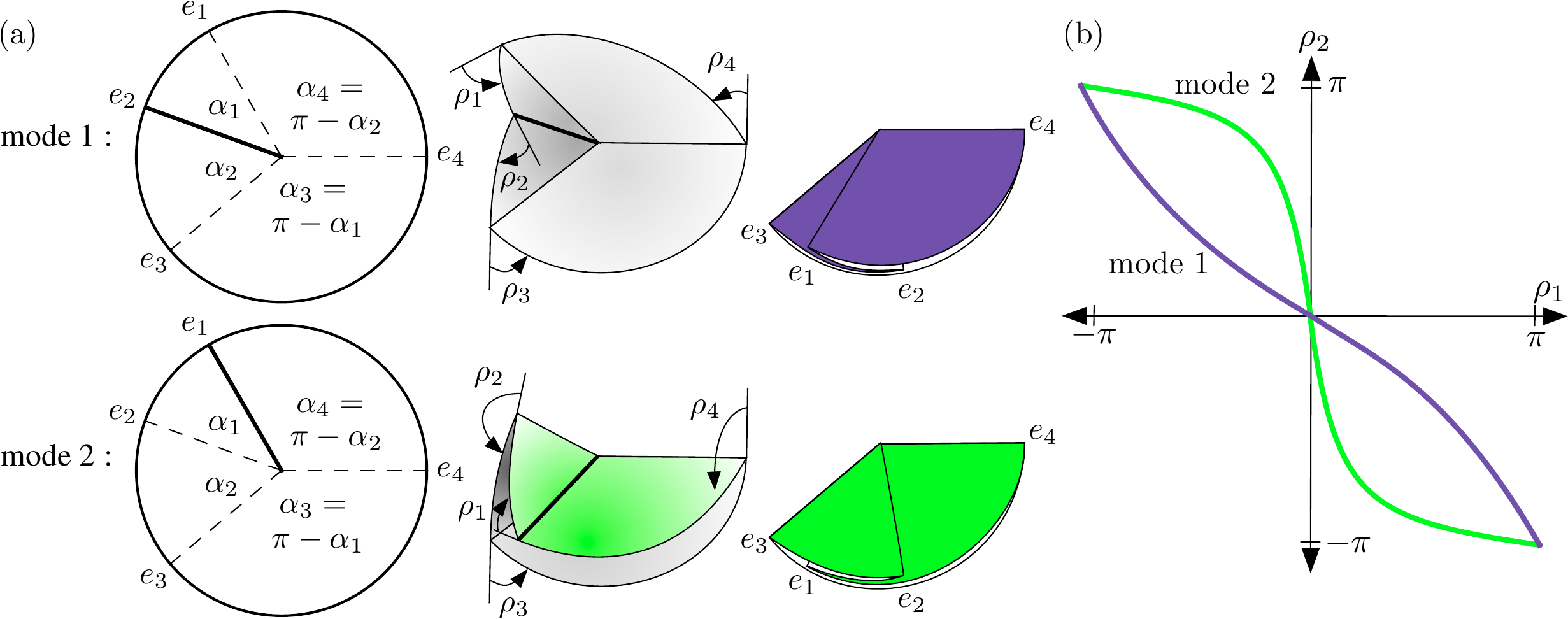}
\caption{(a) The two folding modes of a developable, flat-foldable vertex of degree 4. (b) The configuration space curves for folding angles $\rho_1$ and $\rho_2$.}\label{fig1}
\end{figure*}

When $V$ is a flat-foldable degree-4 origami vertex, we have the following (see Figure~\ref{fig1}(a) to aid in the notation).

\begin{theorem}\label{thm1}
For a developable, flat-foldable origami degree-4 vertex with sector angles labeled so that $\alpha_1\leq~\alpha_2<\alpha_3,~\alpha_4$, the folding angles $\rho_i$ satisfy one of the following sets of equations:
\begin{equation}\label{eq1}
\rho_1=\rho_3, \rho_2=-\rho_4, \tan\frac{\rho_2}{2} = -\frac{\cos\frac{\alpha_1+\alpha_2}{2}}{\cos\frac{\alpha_1-\alpha_2}{2}} \tan\frac{\rho_1}{2}, 
\end{equation}
and
\begin{equation}\label{eq2}
\rho_1=-\rho_3, \rho_2=\rho_4, \tan\frac{\rho_1}{2} = \frac{\sin\frac{\alpha_1-\alpha_2}{2}}{\sin\frac{\alpha_1+\alpha_2}{2}} \tan\frac{\rho_2}{2}.
\end{equation}
\end{theorem}

For a proof, see \cite{Origametry} or \cite{TachiHull}.  The two sets of equations in Theorem~\ref{thm1} trace two curves in the configuration space, called the \textit{modes} of the rigid folding, that intersect at the origin (the unfolded state).  A example of $(\rho_1,\rho_2)$ graphs of these two modes is shown in Figure~\ref{fig1}(b). 

These relationships were first described by Huffman in 1976, although not in this exact form \cite{Huff,Huff2}.  
We see that developable, degree-4 flat-foldable origami vertices have opposite folding angles that are congruent up to sign, and adjacent folding angles have a linear relationship when parameterized by the tangent half-angle.

For the developable, non-flat-foldable case Huffman \cite{Huff} provides a relationship for the opposite folding angles,
\begin{equation}\label{eq3}
\sin^2\frac{\rho_i}{2} = \frac{\sin\alpha_{i+1} \sin\alpha_{i+2}}{\sin\alpha_i \sin\alpha_{i+3}} \sin^2\frac{\rho_{i+2}}{2},
\end{equation}
where the indices are taken cyclically (mod 4), and for the adjacent folding angle relationships  Huffman gives a very convoluted expression, lamenting that a more simple expression does not seem possible \cite[p. 1014]{Huff}.  Izmestiev \cite{Izmestiev} provides formulas that are of similar, but simplified, form to Huffman's. However, they exist in a complexified configuration space and are thus challenging to use.  

Numerical methods have also been employed to calculate folding angles, such as in Tachi's Rigid Origami Simulator software \cite{Tachi2}.  Numerical methods have also been used to compute configuration space curves of non-developable degree-4 vertex rigid foldings, such as those in \cite{Toptrans20,NEO20}, which classify the possible combinations of mountains and valleys that can exist in non-Euclidean vertices.

We now turn our attention to deriving folding angle equations for rigid foldings of degree-4 origami vertices in general. \\

\section{Kinematics of general degree-4 vertices}\label{sec3}

The following Theorem describes the folding angle relationships, and thus the configuration space, of rigid foldings of general degree-4 origami vertices.

\begin{theorem}\label{thm2}
Given a general degree-4 rigid origami vertex with sector angles $\alpha_1,\ldots,\alpha_4$ and creases $e_i$ between sectors $\alpha_{i-1}$ and $\alpha_{i}$, the configuration space of the vertex is the set of folding angles $(\rho_1,\ldots, \rho_4)$ of the creases $e_i$ that satisfy the following two equations, where the subscript index arithmetic is taken cyclically (mod 4):
\begin{widetext}
\begin{equation}\label{confeq1}
\begin{multlined}
\tan^2\frac{\rho_i}{2} = 
-\frac{(1+\tan^2\frac{\rho_{i+2}}{2})\cos(\alpha_{i-1}+\alpha_i)+\tan^2\frac{\rho_{i+2}}{2}\cos(\alpha_{i+1}-\alpha_{i+2})+\cos(\alpha_{i+1}+\alpha_{i+2})}
{(1+\tan^2\frac{\rho_{i+2}}{2})\cos(\alpha_{i-1}-\alpha_i)-\tan^2\frac{\rho_{i+2}}{2}\cos(\alpha_{i+1}-\alpha_{i+2})-\cos(\alpha_{i+1}+\alpha_{i+2})}
\end{multlined}
\end{equation}
and
\begin{equation}\label{confeq2}
\begin{split}
\cos\alpha_{i+2} \left(1+\tan^2\frac{\rho_i}{2}\right)\left(1+\tan^2\frac{\rho_{i+1}}{2}\right) =\quad 
&\cos(\alpha_{i+1}-\alpha_i-\alpha_{i-1}) \tan^2\frac{\rho_{i+1}}{2} \\ 
+ &\cos(\alpha_{i+1}+\alpha_i-\alpha_{i-1}) \tan^2\frac{\rho_i}{2} \\
+ &\cos(\alpha_{i+1}-\alpha_i+\alpha_{i-1}) \tan^2\frac{\rho_i}{2}\tan^2\frac{\rho_{i+1}}{2} \\
+ &\cos(\alpha_{i+1}+\alpha_{i}+\alpha_{i-1}) \\
+ & 4\sin\alpha_{i+1} \sin\alpha_{i-1} \tan\frac{\rho_i}{2}\tan\frac{\rho_{i+1}}{2}.
\end{split}
\end{equation}
\end{widetext}

\end{theorem}

Proofs of these equations can be found in Appendix~\ref{appA}.  

\begin{remark}\label{rem1}
Equation~\eqref{confeq1} describes the relationship between opposite pairs of folding angles at the degree-4 vertex, while Equation~\eqref{confeq2} describes adjacent pairs of folding angles.  These equations capture the entire configuration space of a degree-4 rigidly-folding vertex, but to obtain functions for individual folding angles they need to be manipulated, whereby the choice of square root branches determines the various folding modes.  To enumerate these modes, note that since degree-4 origami vertices have one degree of freedom, we may choose any angle to parameterize the rigid folding.  For example, if we let $t=\rho_4$ be the parameter, then from Equation~\eqref{confeq2} we obtain two solutions for $\rho_1$.  That is, we can isolate the $\tan(\rho_1/2)$ terms to obtain
\begin{widetext}
\begin{equation}\label{confeq3}
\tan \frac{\rho_{1}}{2} = \\
\frac{2\sin\alpha_1 \sin\alpha_3 \tan\frac{\rho_4}{2}\pm
\sqrt{\hspace{-.05in}
\begin{array}{l}
4\sin^2\alpha_{1} \sin^2\alpha_{3} \tan^2\frac{\rho_4}{2} \\
-(\cos\alpha_{2} - \cos(\alpha_{1}-\alpha_{3} -\alpha_4) + (\cos\alpha_{2}  - \cos(\alpha_{1}+\alpha_{3} -\alpha_4))\tan^2\frac{\rho_4}{2})\\
\times (\cos\alpha_{2} - \cos(\alpha_{1}+\alpha_{3} +\alpha_4)+(\cos\alpha_{2} - \cos(\alpha_{1}-\alpha_{3} +\alpha_4))\tan^2\frac{\rho_4}{2})
\end{array}
}    }
{\cos\alpha_{2} - \cos(\alpha_{1}-\alpha_{3} -\alpha_4)+(\cos\alpha_{2}-\cos(\alpha_{1}+\alpha_{3} -\alpha_4))\tan^2\frac{\rho_4}{2}},
\end{equation}
\end{widetext}
giving two choices for $\rho_1$. If we  keep the sector with angle $\alpha_4$ fixed and fold the creases on either side of this sector with the folding angles $\rho_1$ and $\rho_4$, then this will position the sectors $\alpha_1$ and $\alpha_3$, resulting in only one way to place the sector with angle $\alpha_2$ between them, determining the folding angles $\rho_2$ and $\rho_3$ (although this may result in the material self-intersecting).  In other words,  Equation~\eqref{confeq2} provides a proof of the following (which is already generally known, e.g. see \cite{Huff,Waitukaitis16,Izmestiev}):

\begin{corollary}
Degree-4 rigid-foldable vertices have exactly two folding modes, meaning that their configuration space consists of two curves in $\mathbb{R}^4$.
\end{corollary}

\end{remark}

\begin{remark}
Care must be taken when trying to use Equations~\eqref{confeq1} and \eqref{confeq2} to isolate the different folding modes, such as to express one folding angle as a function of another folding angle.  For example, when taking square roots of both sides of Equation~\eqref{confeq1} one needs to track which branch of the square root is needed to preserve the folding mode.   

A guiding principle that can help, especially for non-Euclidean degree-4 vertices, is to choose a folding angle $\rho_i$ as the independent parameter that can achieve the full range of $[-\pi,\pi]$ in the rigid folding motion. For example, consider a \textit{bird's foot} vertex, which is a degree-4 vertex with $\alpha_1=\alpha_2<\alpha_3=\alpha_4$, so that creases $e_1$-$e_3$ look like the toes of a bird's foot (see Figure~\ref{fig2.5}(a) for an example). In the elliptic case one should choose $\rho_2$, the folding angle of crease $e_2$, to be the free parameter, since all the other folding angles only remain in the range $[0,\pi]$ while in mode 1; if we make any of $\rho_1$, $\rho_3$, or $\rho_4$ negative, then we jump to a different connected component of the configuration space and will be in mode 2.  (More details of how the folding angles equations from Theorem~\ref{thm1} inform the rigid folding of bird's feet can be found in Appendix~\ref{appC}.)
\end{remark}

\begin{figure*}
\centerline{\includegraphics[width=\linewidth]{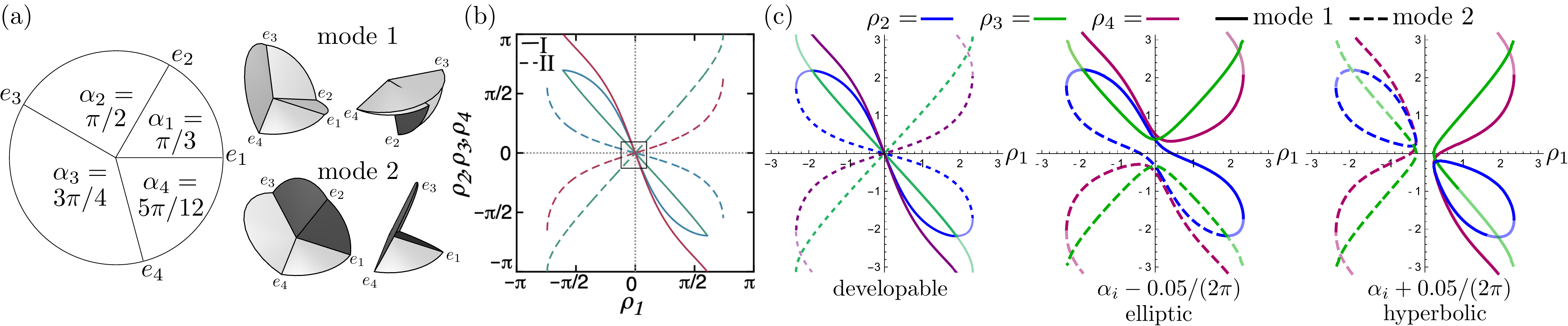}}
\caption{(a) An example of a developable degree-4 origami vertex and images of it rigidly folding in its two folding modes.  (b) Graphs of the folding angle relationships $(\rho_1,\rho_2)$, $(\rho_1,\rho_3)$, and $(\rho_1,\rho_4)$ for this vertex as computed in \cite{NEO20} (reprinted with permission). (c) The folding angle relationships for this vertex as plotted from the equations in Theorem~\ref{thm2} for the Euclidean developable case and perturbations into the non-Euclidean elliptic and hyperbolic cases. Faded parts of curves correspond to self-intersections of the material or folding angles going beyond $\pm\pi$.}\label{fig2}
\end{figure*}

\begin{example}\label{ex1}
Equations \eqref{confeq1} and \eqref{confeq2} can be used to verify many of the qualitative behaviors of degree-4 rigid origami vertices as documented by Waitukaitis et al. for the developable case in \cite{Waitukaitis16} and the non-developable case in \cite{NEO20}.  For example, in \cite{NEO20} the developable degree-4 vertex with plane angles $\alpha_1=\pi/3$, $\alpha_2=\pi/2$, $\alpha_3=3\pi/4$, and $\alpha_4=5\pi/12$ is studied; see Figure~\ref{fig2}(a)-(b) for images of this vertex and a plot of the folding angle relationships between $\rho_1$ and $\rho_2$, $\rho_3$, and $\rho_4$ for both folding modes, from \cite{NEO20} (reprinted with permission).  Using Equations \eqref{confeq1} and \eqref{confeq2} from Theorem~\ref{thm2}, we can plot these curves as shown in the first graph of Figure~\ref{fig2}(c), showing our equations match previous work in the developable case. In the remaining two plots of Figure~\ref{fig2}(c) we show our algebraic configuration space curves for this vertex where the sector angles $\alpha_i$ have been perturbed by $-0.05/(2\pi)$, making a synclastic convex cone vertex, and by $0.05/(2\pi)$ to make an anticlastic hyperbolic vertex. Our curves match the approximation curves made for such non-Euclidean vertices near the origin given in \cite[Fig. 2]{NEO20}. Note that in this Figure we plot all solutions of the Theorem~\ref{thm2} equations that give $\rho_i\in [\pi,\pi]$, which includes rigid foldings that cause the material to self-intersect or cause other folding angles to go beyond $\pm\pi$; such solution curve parts are drawn in faded line widths in Figure~\ref{fig2}(c).  

Many things can be inferred from the curves in Figure~\ref{fig2}(c).  For one thing, in both the elliptic and hyperbolic cases the configuration spaces are disconnected, whereas in the developable vertex they are connected.  This makes intuitive sense because when $\sum\alpha_i\ne 2\pi$, it is impossible to unfold the vertex so that all the creases have folding angles of zero at the same time. In the elliptic case this means that the vertex can ``pop up" or ``down" and cannot switch between the two without bending faces.  It is less intuitive that one cannot switch between modes 1 and 2 in the hyperbolic case, yet their curves clearly do not intersect.  This topological distinction of the configuration spaces in the elliptic, developable, and hyperbolic cases has further implications on the definitions of modes 1 and 2.
E.g., in the developable vertex the folding angles $\rho_3$ and $\rho_4$ are decreasing functions of $\rho_1$ in mode 1 passing from valley, to unfolded (zero), to mountain, ensuring a smooth folding motion. But in the elliptic case $\rho_3$ and $\rho_4$ remain valleys throughout mode 1, joining two branches that were in different branches in the developable case.  Such observations could be useful in practice when, say, trying to decide which crease to use to drive a rigid folding mechanism.
\end{example}

\section{General equations for developable degree-4 vertices}\label{sec4}

Despite the elegant folding angle equations for flat-foldable degree-4 vertices, as shown in Theorem~\ref{thm1}, equally elegant equations for general degree-4 developable vertices have been elusive in the literature.  However, the equations in Theorem~\ref{thm2} can be used to prove (see Appendix~\ref{appB}) the following:

\begin{theorem}\label{thm3}
For a general, developable degree-4 rigid origami vertex with sector angles $\alpha_i$ and folding creases $e_i$ between sectors $\alpha_{i-1}$ and $\alpha_i$, the folding angles $\rho_i$ at $e_i$ satisfy the following equations:
\begin{equation}\label{gdeq1}
\begin{split}
 \frac{\sin\alpha_{i+1} \sin\alpha_{i+2}}{\tan^2\frac{\rho_i}{2}} &+ \frac{\cos(\alpha_{i+1}-\alpha_{i+2})}{2} = \\
 \frac{\sin\alpha_{i-1}\sin\alpha_i}{\tan^2\frac{\rho_{i+2}}{2}} &+ \frac{\cos(\alpha_{i-1}-\alpha_i)}{2}
\end{split}
\end{equation}
and
\begin{equation}\label{gdeq2}
\frac{\sin(\alpha_{i+1}+\alpha_{i+2})}{\tan\frac{\rho_i}{2}} = 
\frac{\sin\alpha_i}{\tan\frac{\rho_{i-1}}{2}} + \frac{\sin\alpha_{i-1}}{\tan\frac{\rho_{i+1}}{2}}.
\end{equation}
\end{theorem}

\begin{remark}
Equation~\eqref{gdeq1} can be manipulated to become Huffman's Equation~\eqref{eq3} from \cite{Huff}, but the formulation in \eqref{gdeq1} shows how the general equation for opposite folding angles can be expressed with tangent of half the folding angles.

Equation~\eqref{gdeq2} reveals a pattern that was hidden in the flat-foldable equations in Theorem~\ref{thm1}.  Since in the flat-foldable case we have $\rho_{i-1} = \pm \rho_{i+1}$, the right-hand side of Equation~\eqref{gdeq2} becomes either
$$\frac{\sin\alpha_i + \sin\alpha_{i-1}}{\tan\frac{\rho_{i-1}}{2}}\mbox{ or }\frac{\sin\alpha_i - \sin\alpha_{i-1}}{\tan\frac{\rho_{i-1}}{2}},$$
which, using $\alpha_{i+1} = \pi-\alpha_{i-1}$ and $\alpha_{i+2}=\pi-\alpha_i$, makes Equation~\eqref{gdeq2} one of the two folding modes of Theorem~\ref{thm1} (since, for example, $\sin(\alpha_i+\alpha_{i-1})/(\sin\alpha_i + \sin\alpha_{i-1}) = \cos((\alpha_i+\alpha_{i-1})/2)/\cos((\alpha_i-\alpha_{i-1})/2)$). Thus, the flat-foldable case collapses the folding angles $\rho_{i-1}, \rho_{i+1}$ of the two opposing creases to relate them to $\rho_i$, whereas Equation~\eqref{gdeq2} shows how they should be separated in the general, developable case.
\end{remark}

\section{Applications}\label{sec5}

We now illustrate how the equations from Theorem~\ref{thm2} can be used to analyze specific examples of rigid-foldable crease patterns and explain their mechanical behavior.

\subsection{Square twist pouches}

\begin{figure}
\centerline{\includegraphics[width=\linewidth]{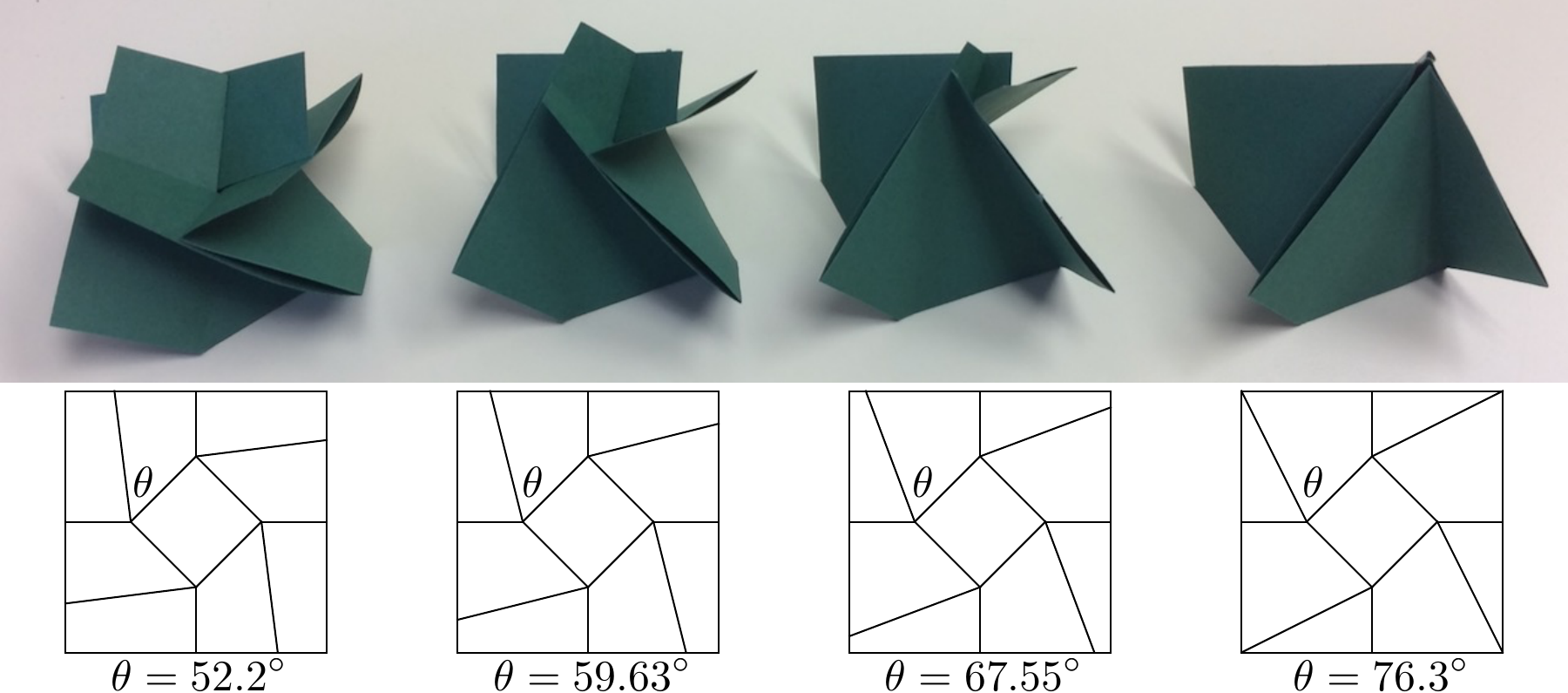}}
\caption{Square twist crease pattern variations and their folded results.  If $\theta=45^\circ$ this is the classic square twist.  For   $\theta>45^\circ$ the vertices become non-flat-foldable and fold to a pouch-like non-flat state.}\label{fig3}
\end{figure}

A \textit{square twist} is a flat-foldable crease pattern made of four degree-4 vertices forming a square in the paper, whereby flat-folding all the creases causes this square of paper to rotate, or twist.  Square twists have been studied extensively for their bistable properties and as a building block for larger origami mechanisms \cite{Silverberg2,EvansLang2}.  For instance, if the creases of the square connecting the four vertices are made to be all valleys (or all mountains), then the square twist will have only two rigidly-folded states, the unfolded state and the flat-folded state where all folding angles are $\pm\pi$.

\begin{figure}
\centerline{\includegraphics[scale=.25]{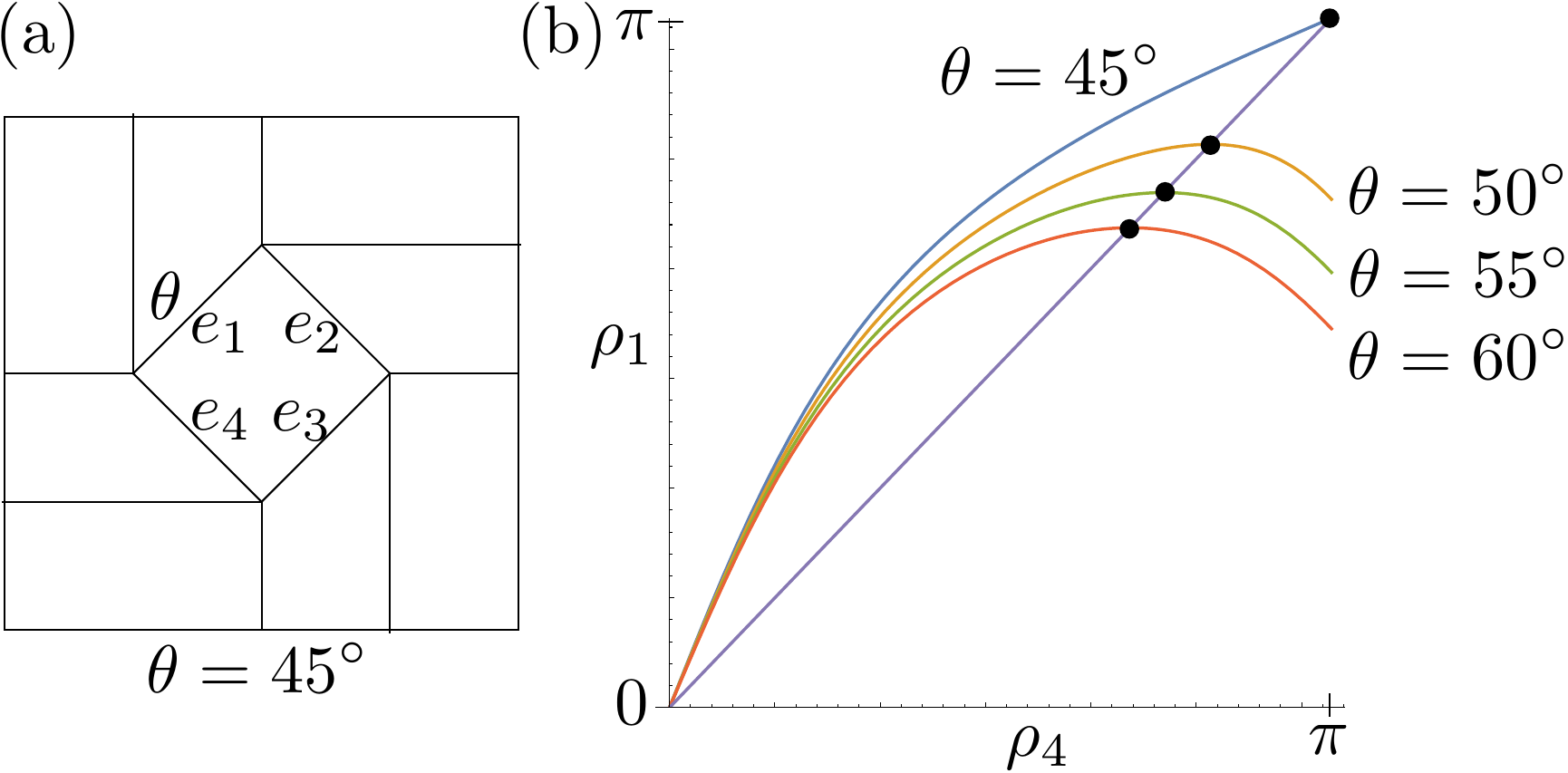}}
\caption{(a) The square twist crease pattern with creases $e_1$ thru $e_4$ labeled. (b) Plots of the $(\rho_4,\rho_1)$ equation derived from Theorem~\ref{thm2} for various $\theta$, along with the $\rho_4=\rho_1$ line, with their fixed points indicated.}\label{fig4}
\end{figure}

The classic, flat-foldable square twist crease pattern is as shown in Figure~\ref{fig4}(a) with $\theta=45^\circ$.  If we increase $\theta$ at all four vertices (so that the crease pattern is still rotationally symmetric) then the vertices become non-flat-foldable and the crease pattern, when making the inner square be all valleys, will form a 3D pouch when folded.  Examples for various $\theta>45^\circ$ are shown in Figure~\ref{fig3}.  Such origami pouches have been explored by a number of origami artists and researchers such as Chris Palmer \cite{Palmer} and Jun Mitani \cite{Mitani}.
Like the flat-foldable square twist, these square twist pouches have only two rigidly-foldable states, and when folding these crease patterns physically one can feel the paper ``snap" into the rigid folded state.  That is, these crease patterns exhibit bistability (between the unfolded state and a unique rigid-folded state) like the flat-folded studies in \cite{Silverberg2}.

To prove the bistabiity of these square twist pouches we can plot Equation~\eqref{confeq3} with $\alpha_1=\theta=45^\circ + \Delta$, $\alpha_2=90^\circ-\Delta$, $\alpha_3=135^\circ$, and $\alpha_4=90^\circ$ for various values of $\theta$, with $0\leq\theta\leq 90^\circ$.  Such plots on a $(\rho_4, \rho_1)$ axis are shown in Figure~\ref{fig4}(b).  Where these plots cross the $\rho_4=\rho_1$ line represent configurations for this degree-4 vertex that have equal folding angles at creases $e_1$ and $e_4$ (see Figure~\ref{fig4}(a)). This is the only case that will allow non-zero folding angles to be used at each vertex to rigidly fold the whole crease pattern and maintain a consistent folding angle loop condition (i.e., rotational symmetry) around the square $e_1$-$e_4$.  One can use spherical trigonometry to prove that the folding angle that allows $\rho_1=\rho_4$ is $\pi-\arccos(\cot\theta)$.  Therefore the rigid origami configuration space of the square twist pouch determined by sector angle $\theta$ is discrete, consisting of only the unfolded state and the two states where the folding angles around the square are all equal to $\pi-\arccos(\cot\theta)$ or its negative.

\subsection{A non-Euclidean folding table}\label{table}

As described in \cite{Toptrans20,NEO20}, the folding mechanics of non-developable degree-4 vertices can be significantly different from developable vertices. In particular non-developable vertices cannot fold to a state where all the folding angles are zero, and therefore, as seen in Example~\ref{ex1} of Section~\ref{sec3}, the configuration spaces for the two folding modes are disconnected. In the hyperbolic  case sometimes (but not always \footnote{Hyperbolic vertices may have non-intersecting states for both modes.  An example can be seen in the hyperbolic bird's foot vertex Figure~\ref{fig2.5}(b), where the mode 2 versions shown are just the mode 1 cases with the Ms and Vs reversed. Whether or not self-intersections will happen within a folding mode depends on subtle differences between the sector and folding angles of the hyperbolic vertex.}) one of the folding modes generated by the Theorem~\ref{thm2} equations will result in the paper self-intersecting.  Such cases can be leveraged in applications, since the self-intersections would make one of the folding modes impossible, guaranteeing a single way to fold the mechanism.  These features are attractive in furniture and architecture designs that employ folding, where controllability (one degree of freedom) and consistency (one folding mode) are essential.

\begin{figure}
\centerline{\includegraphics[scale=.25]{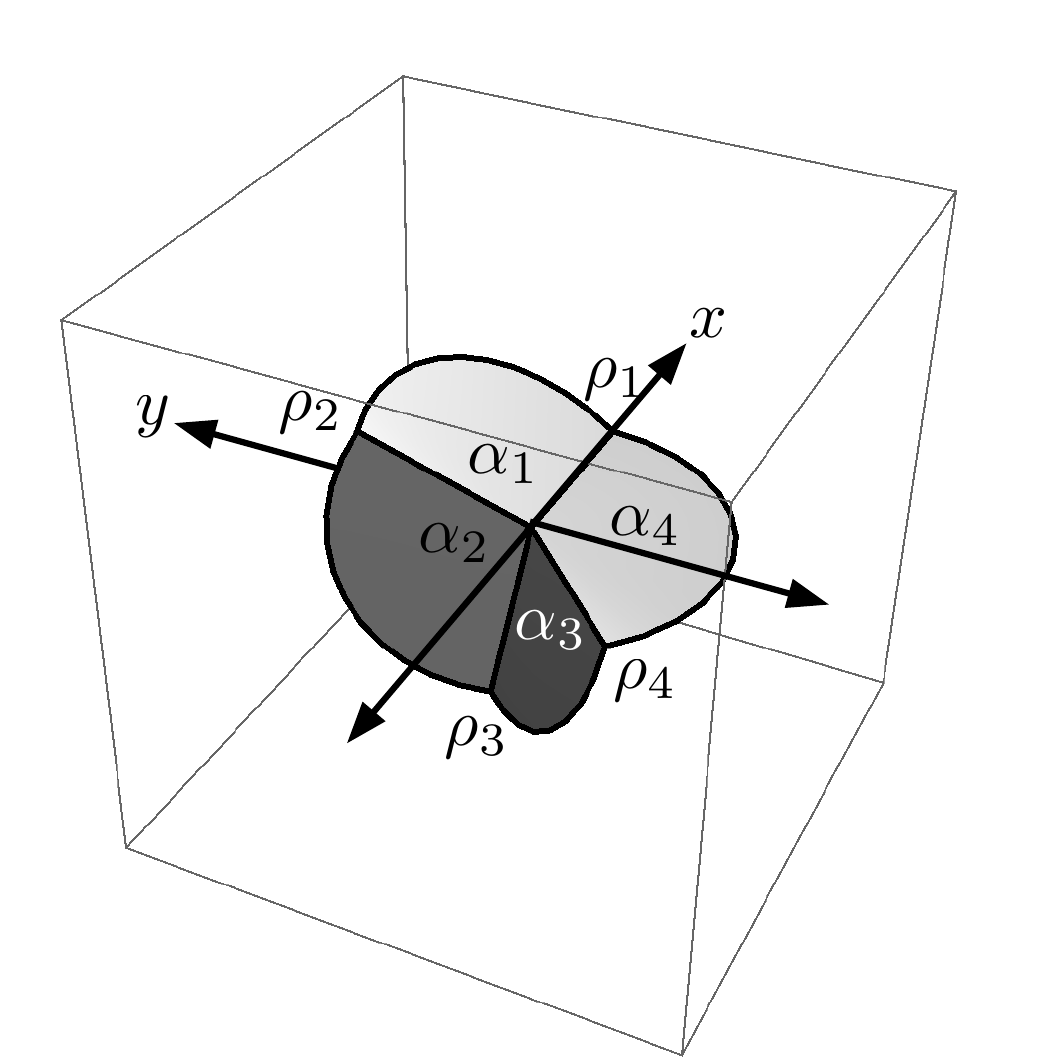}}
\caption{A rigid folding of a non-developable, hyperbolic vertex with $\alpha_1=\alpha_2=5\pi/8$, $\alpha_3=\pi/2$, and $\alpha_4=3\pi/4$.}\label{fig4.5}
\end{figure}

\begin{figure} 
    \centering
    \includegraphics[width=7cm]{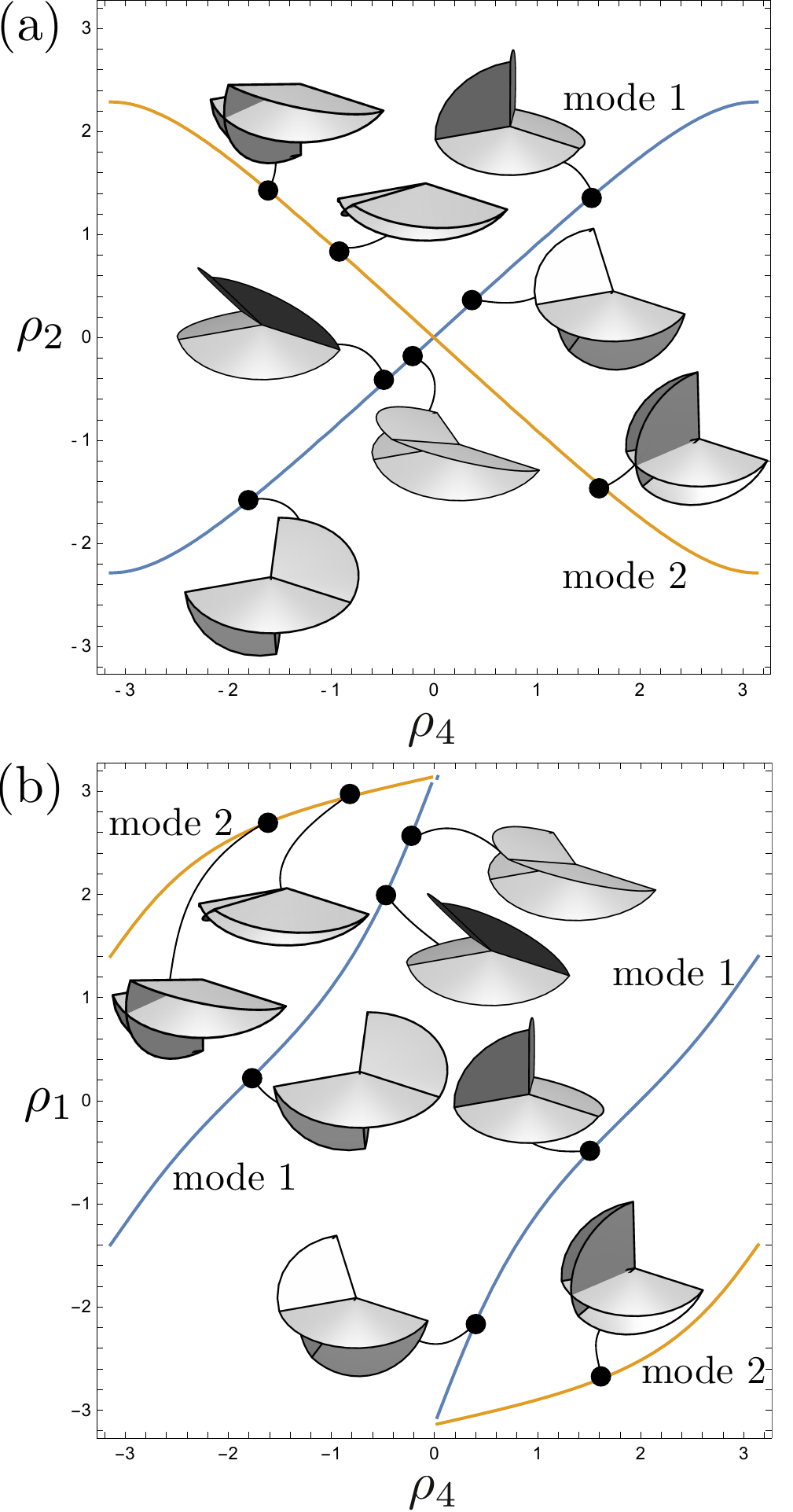}
    \caption{(a) The configuration space of $\rho_2$ graphed with $\rho_4$, following Equation~\eqref{teq1}, with some points indicated with their rigid foldings. (b) The curves for $\rho_1$ graphed with $\rho_4$ using Equation~\eqref{confeq2} and the same rigid folding sample points.}\label{fig5}
\end{figure}

An example is a folding table designed by author Foschi whose crease pattern is made of eight degree-4 hyperbolic vertices, all with sector angles $\alpha_1=\alpha_2=5\pi/8$, $\alpha_3=\pi/2$, and $\alpha_4=3\pi/4$.  Such a vertex is shown in Figure~\ref{fig4.5}.  Entering these into Equation~\eqref{confeq1} and simplifying reveals
\begin{equation}\label{teq1}
\tan^2\frac{\rho_2}{2} = \frac{2\sqrt{2}}{2+\sqrt{2}\cos \rho_4}\sin^2\frac{\rho_4}{2}.
\end{equation}
The two folding modes can be separated from Equation~\eqref{teq1} as 
$$\tan\frac{\rho_2}{2}=\left\{
\begin{array}{cl}
\sqrt{2\sqrt{2}/(2+\sqrt{2}\cos \rho_4)}\sin\frac{\rho_4}{2} & \mbox{for mode 1,}\\ 
-\sqrt{2\sqrt{2}/(2+\sqrt{2}\cos \rho_4)}\sin\frac{\rho_4}{2} & \mbox{for mode 2.}
\end{array}\right.$$ 
Similarly, entering our sector angles $\alpha_i$ into Equation~\eqref{confeq3} and simplifying produces the corresponding mode folding angle equations for $\rho_1$:
$$\tan\frac{\rho_1}{2}=\left\{
\begin{array}{cl}
\frac{2+2\sqrt{2}-\sqrt[4]{2^3}\sqrt{3+2\sqrt{2}+\tan^2\frac{\rho_4}{2}}}{2\tan\frac{\rho_4}{2}} & \mbox{for mode 1,}\\ 
\frac{2+2\sqrt{2}+\sqrt[4]{2^3}\sqrt{3+2\sqrt{2}+\tan^2\frac{\rho_4}{2}}}{2\tan\frac{\rho_4}{2}} & \mbox{for mode 2.}
\end{array}\right.$$ 
Graphs of the mode curves for $(\rho_4, \rho_2)$ and $(\rho_4,\rho_1)$ are shown in Figure~\ref{fig5} along with images of the folded vertex at a few points.  This Figure also suggests that the vertex is a valid rigid fold in mode 1 but self-intersects in mode 2.  This can be verified by noting that, according to the graphs of our folding angle equations, for mode 1 we have an alternating MV assignment MVMV whereas for mode 2 we have MMVV.  The latter cannot be folded rigidly with this collection of sector angles without forcing a self-intersection. (See \cite{NEO20} for a detailed description of which MV combinations can be achieved in non-Euclidean vertices.)  Note that hyperbolic vertices do not always force self-intersections in one of their folding modes; the discussion of bird's feet vertices in Appendix~\ref{appC} shown one example.

Therefore the non-developable vertex with sector angles $(5\pi/8, 5\pi/8, \pi/2, 3\pi/4)$ has only one physically foldable folding mode, which is mode 1 in Figure~\ref{fig5}.  Note that the origin of Figure~\ref{fig5}(a) represents the folding where $\rho_2 = \rho_4=0$ and we have a double-covered flat fold.  Also, comparing the graphs in Figure~\ref{fig5}(a) and (b), we see that as $\rho_4$ approaches 0 in the negative direction in mode 1, we will have $\rho_1\rightarrow \pi$, whereas if $\rho_4$ approaches 0 in the positive direction in mode 1 we have $\rho_1\rightarrow -\pi$.  Thus $\rho_4=0$ is a discontinuity for the $\rho_1$ (and $\rho_3$) folding angle, as can be seen in Figure~\ref{fig5}(b).  In other words, in a physical model we cannot fold from negative $\rho_4$ values to positive $\rho_4$, since doing so would cause the crease $e_1$ to turn from a valley to a mountain as we pass through the origin, and this would require the folded material to pass through itself.

\begin{figure} 
    \centering
    \includegraphics[width=\linewidth]{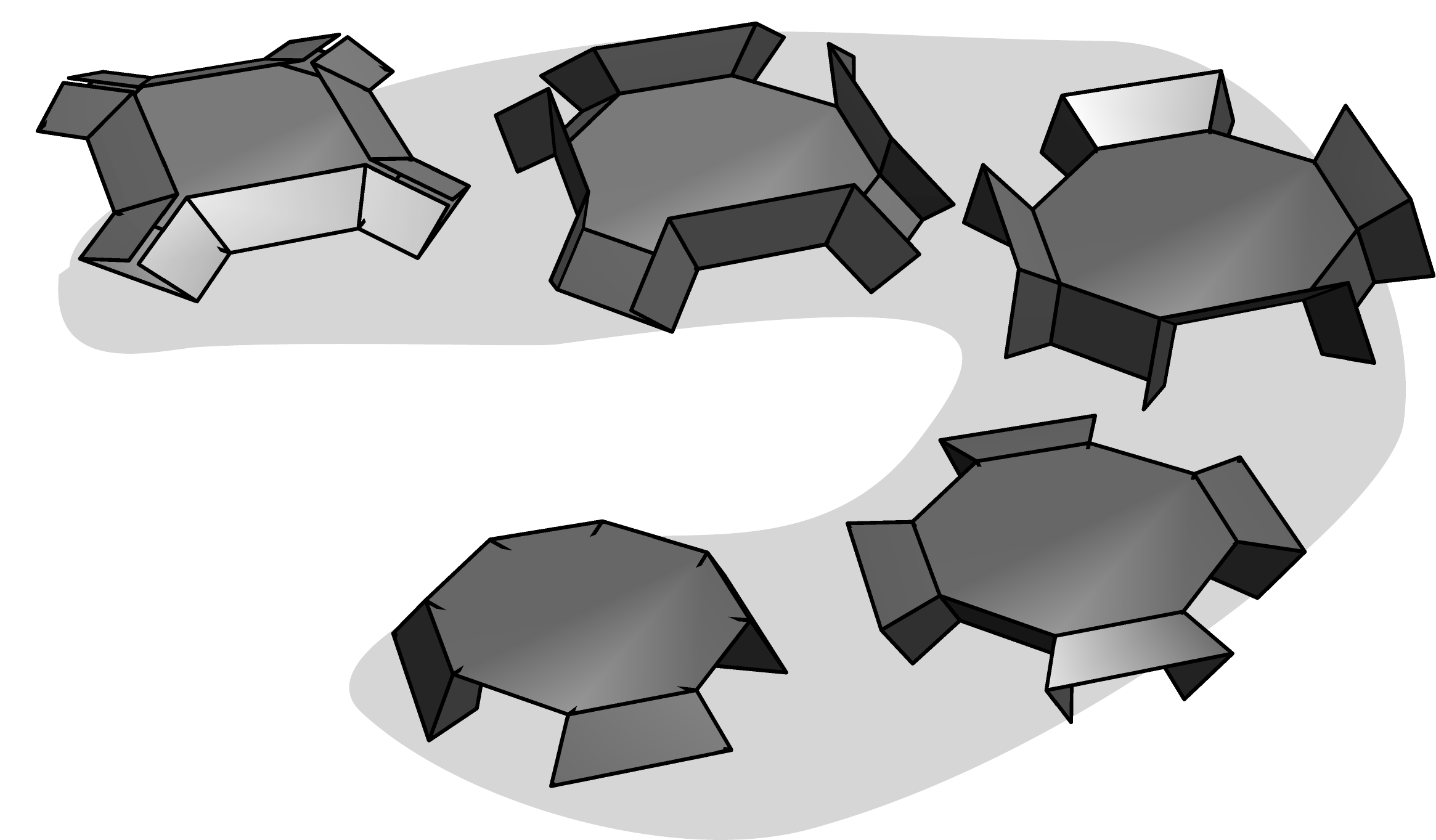}
    \caption{The complete non-developable folding table.}\label{fig6}
\end{figure}

When the non-developable $(5\pi/8, 5\pi/8, \pi/2, 3\pi/4)$ vertices are placed together to make an octagonal ring, the non-developable folding table is formed, as shown in Figure~\ref{fig6}.  Note that the construction of the table has the layers of material arranged from the start so that the creases labeled $e_4$ in our single-vertex analysis must fold into mountain creases, implying that $\rho_4<0$ throughout the folding process.  In other words, the rigid folding motion shown in Figure~\ref{fig6} is the only valid folding motion for this table, leading to a mechanical folding design that cannot misfold into an undesired shape.

\section{Conclusion}

We have devised folding angle equations that hold for all cases of degree-4 rigid origami vertices.  While it was previously known that such equations exist, the ones presented here have the advantage of being expressed in terms of tangents of half the folding angles, which allows us to make connections to the flat-foldable case. In addition, we used these equations to prove surprisingly simple folding angle equations for arbitrary degree-4 vertices in the developable case which nicely generalize the flat-foldable case.  Also, we provided some examples of how these equations can easily help analyze the kinematic behavior of degree-4 rigid origami vertex designs.

The fact that the tangent half-angle representation is so prevalent in degree-4 folding angle equations remains to be fully understood.  As detailed in \cite{FHR22}, sometimes this phenomenon can be explained by proving that a given crease pattern is kinematically equivalent to a developable flat-foldable crease pattern, where the folding angle equations are linear in terms of $\tan(\rho_i/2)$. It could be that this technique can provide a different proof of the general degree-4 equations in Theorem~\ref{thm3}, but it is not clear how this would be done for non-developable degree-4 vertices.

\begin{acknowledgments}
The authors would like to thank Robert J. Lang and Tomohiro Tachi for helpful conversations. Part of this project grew out of the 2022 Structural Origami Gathering workshop organized by Tomohiro Tachi and founded by Rupert Maleczek. Author T.C.H is supported by NSF grant DMS-1906202.
\end{acknowledgments}

\appendix

\section{Proof of the general degree-4 equations}\label{appA}

\begin{figure} 
    \centering
    \includegraphics[scale=.4]{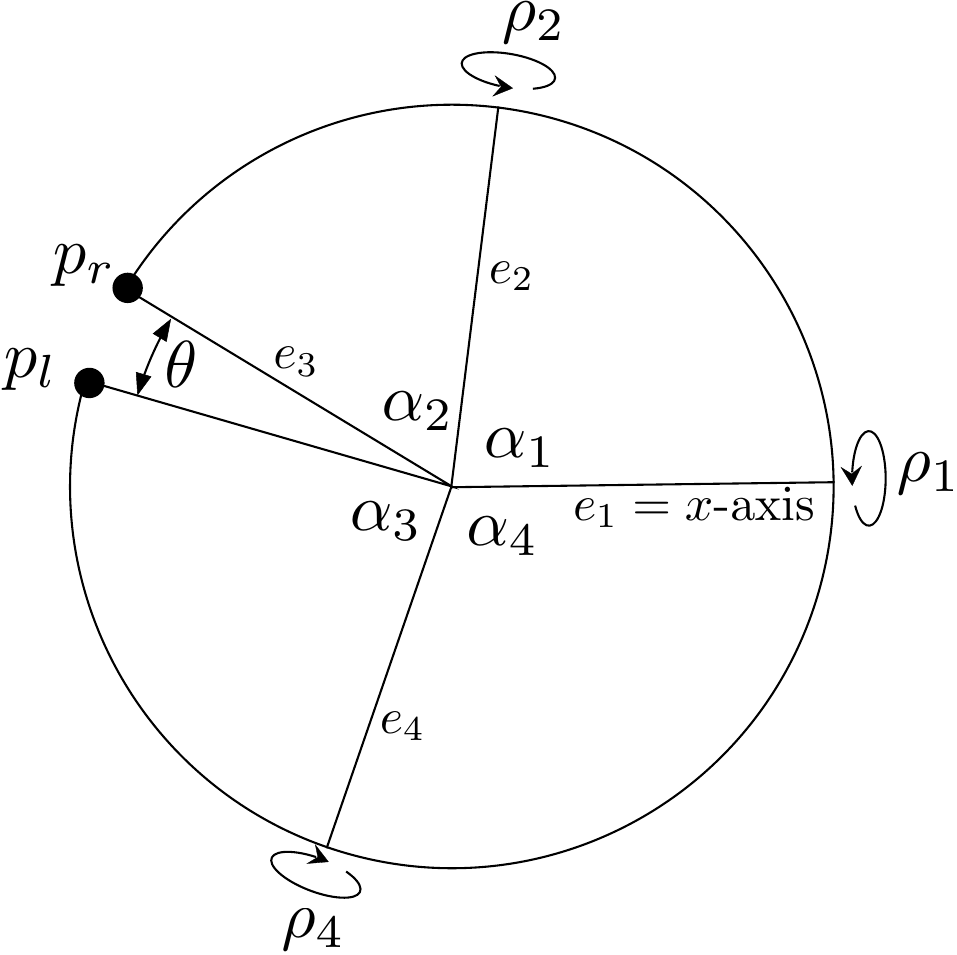}
    \caption{A degree-4 vertex crease pattern.  If $\theta=0$ the vertex is developable; $\theta\ne 0$ is non-developable.}\label{afig1}
\end{figure}

A schematic of an arbitrary degree-4 vertex is shown in Figure~\ref{afig1}, which we consider lying in the $xy$-plane with the vertex at the origin and the crease $e_1$ along the positive $x$-axis (note that rigid foldability requires $0<\alpha_i<\pi$ for each $\alpha_i$).  In order to make the vertex lie flat in the $xy$-plane, and to aid our kinematic analysis, we split crease line $e_3$ and insert an angle $\theta$.  If $\theta=0$ then the vertex is developable ($\sum \alpha_i= 2\pi$).  The non-developable cases are $\theta>0$, where the vertex forms an elliptic polyhedral cone, and $\theta<0$ which gives us a hyperbolic vertex.

We will follow an approach to modeling the kinematics of all these cases based on rotation matrices (e.g.,  \cite{Devin,Devin2}).  We let the sector of paper with angle $\alpha_4$ to remain fixed  and let $\rho_4 = t$ be the free parameter.  This means that the point $p_l = (\cos(2\pi-\alpha_3-\alpha_4), \sin(2\pi-\alpha_3-\alpha_4))$ will fold to position
\begin{equation}\label{aeq1}
R_z(-\alpha_4)R_x(t)R_z(\alpha_4)p_l
\end{equation}
where $R_x(\beta)$ and $R_z(\beta)$ are the $3\times 3$ matrices that rotate $\mathbb{R}^3$ by $\beta$ about the $x$- and $z$-axes, respectively.
On the other hand the point $p_r=(\cos(\alpha_1+\alpha_2), \sin(\alpha_1+\alpha_2))$ will fold into position
\begin{equation}\label{aeq2}
R_x(\rho_1)R_z(\alpha_1)R_x(\rho_2)R_z(-\alpha_1)p_r.
\end{equation}
The $x$-coordinate of \eqref{aeq2} does not involve $\rho_1$ because the matrix $R_x(\rho_1)$, which leaves the $x$-coordinate invariant, is the only part of \eqref{aeq2} that involves $\rho_1$.  Therefore we can equate the $x$-coordinates of \eqref{aeq1} and \eqref{aeq2} and solve for $\rho_2$.   Doing this gives us
\begin{equation}\label{aeq3}
\cos\rho_2 = \cot\alpha_1\cot\alpha_2 + \frac{\sin\alpha_2\sin\alpha_4\cos t - \cos\alpha_3\cos\alpha_4}{\sin\alpha_1\sin\alpha_2}
\end{equation}

Since cosine is an even function and $\rho_2\in[-\pi,\pi]$, Equation~\eqref{aeq3} implies that there are two possibilities for $\rho_2$ for a given input of $\rho_4=t$, and thus there are at least two folding modes for the vertex which we could denote by $\pm\rho_2(t)$.  An exception for this is when $\alpha_3=\alpha_2$ and $\alpha_4=\alpha_1$, in which case Equation~\eqref{aeq3} reduces to $\rho_2=t$ and there is only one solution for $\rho_2$.

The folding angle $\rho_1$ may then be determined by finding the angle between the vector in \eqref{aeq2} and $R_z(\alpha_1)R_x(\rho_2)R_z(-\alpha_1)p_r$, which is just \eqref{aeq2} with the $R_x(\rho_1)$ removed.  This becomes
\begin{widetext}
\begin{equation}\label{aeq4}
\begin{split}
\rho_1 = &   \arctan \left( \frac{(\cos\alpha_4\sin(\alpha_3+\alpha_4)- \sin\alpha_4 \cos(\alpha_3+\alpha_4))\sin t}
{\cos(\alpha_3+\alpha_4)\sin\alpha_4\cos\alpha_4(\cos t - 1)-\sin(\alpha_3+\alpha_4)(\sin^2\alpha_4 + \cos^2\alpha_4\cos t)}\right) \\
& -\arctan\left( \frac{(\sin(\alpha_1+\alpha_2)- \cos(\alpha_1+\alpha_2)\sin\alpha_1)\sin\rho_2 }
{\cos(\alpha_1+\alpha_2)\cos\alpha_1 \sin\alpha_1 (1-\cos\rho_2) + \sin(\alpha_1+\alpha_2)(\cos^2\alpha_1 \cos\rho_2+\sin^2\alpha_1)}\right)
\end{split}
\end{equation}
\end{widetext}
However, this equation does not lend itself to simplification.  Another approach for relating $\rho_1$ and $\rho_4=t$ is to, keeping the sector $\alpha_4$ in the $xy$-plane fixed, fold $e_1$ and $e_4$ and compute the images of $e_2$ and $e_3$ from this.  The trajectory of $e_2=(\cos\alpha_1,\sin\alpha_1,0)$ and $e_3=(\cos(-(\alpha_3+\alpha_4)),\sin(-(\alpha_3+\alpha_4)),0)$ is
$$T_1:=R_x(\rho_1)e_2\mbox{ and }T_2:=R_z(-\alpha_4)R_x(-\rho_4)R_z(\alpha_4)e_3,$$
respectively.
We want the angle between $T_1$ and $T_2$ to be $\alpha_2$, so $T_1\cdot T_2=\cos\alpha_2$. Expanding this, dividing both sides by $\cos\rho_1 \cos\rho_4$ and rearranging yields
\begin{equation}\label{aeq5}
\begin{split}
\frac{\cos\alpha_1 \cos\alpha_3 \cos\alpha_4 - \cos\alpha_2}{\cos\rho_1 \cos\rho_4} = 
\sin\alpha_1 \sin\alpha_3 \cos\alpha_4 \\
+ \frac{\sin\alpha_1 \cos\alpha_3 \sin\alpha_4}{\cos\rho_4} 
+ \frac{\cos\alpha_1 \sin\alpha_3 \sin\alpha_4}{\cos\rho_1} \\
- \sin\alpha_1\sin\alpha_3\tan\rho_1 \tan\rho_4.
\end{split}
\end{equation}
We then perform the Weierstrass substitution $\sin \rho_4=2x/(1+x^2)$ and $\cos \rho_4 = (1-x^2)/(1+x^2)$, giving us that $x=\tan(\rho_4/2)$.  Then to express $\rho_1$ as a function of $\rho_4$, we also substitute $\sin\rho_1=2y/(1+y^2)$ and $\cos\rho_1 = (1-y^2)/(1+y^2)$, which means $y=\tan(\rho_1/2)$. Substituting these into Equation~\eqref{aeq5} and simplifying gives us
\begin{equation}\label{aeq6}
\begin{split}
\cos(\alpha_1+\alpha_3+\alpha_4) + x^2\cos(\alpha_1-\alpha_3-\alpha_4) \\ + x^2y^2\cos(\alpha_1+\alpha_3-\alpha_4) 
+y^2\cos(\alpha_1-\alpha_3+\alpha_4) \\ + 4xy\sin\alpha_1\sin\alpha_3 -(1+x^2)(1+y^2)\cos\alpha_2=0.
\end{split}
\end{equation}
Re-substituting $x=\tan(\rho_4/2)$ and $y=\tan(\rho_1/2)$ gives us exactly Equation~\eqref{confeq2} from Theorem~\ref{thm2}.




Equation~\eqref{aeq3} above may also be improved by a Weierstrass substitution.  Here we let $\sin t=2x/(1+x^2)$ and $\cos t = (1-x^2)/(1+x^2)$ and $\sin\rho_2=2y/(1+y^2)$ and $\cos\rho_2 = (1-y^2)/(1+y^2)$.  Substituting these into the $x$-coordinates of \eqref{aeq1} and \eqref{aeq2} and isolating the $\rho_2$ terms gives us Equation~\eqref{confeq1}, which captures both folding modes for $\rho_2$.

Since the choice of placing $e_1$ on the positive $x$-axis in this derivation was arbitrary, we could rotate the vertex to place $e_2$, $e_3$, or $e_4$ on the $x$-axis and create similar equations.  This proves Theorem~\ref{thm2}.


\section{Proof of Theorem~\ref{thm3}}\label{appB}

Proving Theorem~\ref{thm3} is a matter of performing extensive trigonometric manipulations to the equations in Theorem~\ref{thm2} along with the fact that, since we're in the developable case, we have $\sum\alpha_i = 2\pi$. We provide an outline of the manipulations needed for the interested reader.

To prove Equation~\eqref{gdeq1}, we use Equation~\eqref{confeq1} from Theorem~\ref{thm2}, where we let $\alpha_{i+2}=2\pi-\alpha_{i-1}-\alpha_i - \alpha_{i+1}$ and simplify to obtain

\begin{widetext}
\begin{equation*}\label{beq1}
\tan^2\frac{\rho_i}{2} = 
\frac{2\sin\alpha_{i+1}\sin(\alpha_{i-1}+\alpha_i+\alpha_{i+1})\tan^2\frac{\rho_{i+2}}{2}}
{\cos(\alpha_{i-1}+\alpha_i)-\cos(\alpha_{i-1}-\alpha_i) + (\cos(\alpha_{i-1}+\alpha_i+2\alpha_{i+1})-\cos(\alpha_{i-1}-\alpha_i))\tan^2\frac{\rho_{i+2}}{2}}.
\end{equation*}
\end{widetext}

Reciprocating, letting $\alpha_{i-1} + \alpha_i + \alpha_{i+1} = 2\pi-\alpha_{i+2}$ again, and separating fractions yields
\begin{equation*}\label{beq2}
\begin{multlined}
\frac{1}{\tan^2\frac{\rho_i}{2}} = 
\frac{\cos(\alpha_{i-1} - \alpha_i) - \cos(\alpha_{i-1}+\alpha_i)}
{2\sin\alpha_{i+1}\sin\alpha_{i+2} \tan^2\frac{\rho_{i+2}}{2}} \\ - 
\frac{\cos(\alpha_{i+1}-\alpha_{i+2})-\cos(\alpha_{i-1}-\alpha_i)}
{2\sin\alpha_{i+1}\sin\alpha_{i+2}}.
\end{multlined}
\end{equation*}
Then using the identity $\cos(a-b)-\cos(a+b) = 2\sin a\sin b$ and simplifying gives us Equation~\eqref{confeq1}.

To verify Equation~\eqref{confeq2} we can start with this equation, take the square root of Equation~\eqref{confeq1} to replace $\tan(\rho_i/2)$, and use Equation~\eqref{confeq3} to replace $\tan(\rho_{i+1}/2)$, making sure to take the square root branches that correspond to the same folding mode (e.g., the positive branch of Eq.~\eqref{confeq1} goes with the negative branch of Eq.~\eqref{confeq3}).  Simlifying this, as well as letting $\alpha_{i+2} = 2\pi - \alpha_{i-1} - \alpha_i - \alpha_{i+1}$, gives us
\begin{equation*}
\begin{split}
\cos\rho_{i+2}\sin\alpha_{i+1}\sin(\alpha_{i-1}+\alpha_i+\alpha_{i+1}) \\ +\cos\alpha_{i+1}\cos(\alpha_{i-1}+\alpha_i+\alpha_{i+1})- \cos(\alpha_{i-1}-\alpha_i) = \\
(1+\cos\rho_{i+2})(\sin\alpha_{i+1}\sin(\alpha_{i-1}+\alpha_i+\alpha_{i-1}) \\ -\sin(\alpha_{i-1}+\alpha_{i+1})\sin(\alpha_i + \alpha_{i+1})(1+\tan^2\frac{\rho_{i+2}}{2}).
\end{split}
\end{equation*}
This equation may then be shown to be true using the identities $\cos\alpha_{i+1}\cos(\alpha_{i-1}+\alpha_i+\alpha_{i+1})- \cos(\alpha_{i-1}-\alpha_i)$ $=$ $2\sin(\alpha_{i-1}+\alpha_{i+1})\sin(\alpha_i+\alpha_{i+1})-\sin\alpha_{i+1}\sin(\alpha_{i-1}+\alpha_i+\alpha_{i+1})$ and $1+\cos\rho_{i+2} = 2\cos^2(\rho_{i+2}/2)$ and simplifying.

\section{Bird's foot vertex examples}\label{appC}

\begin{figure}
    \centering
    \includegraphics[width=\linewidth]{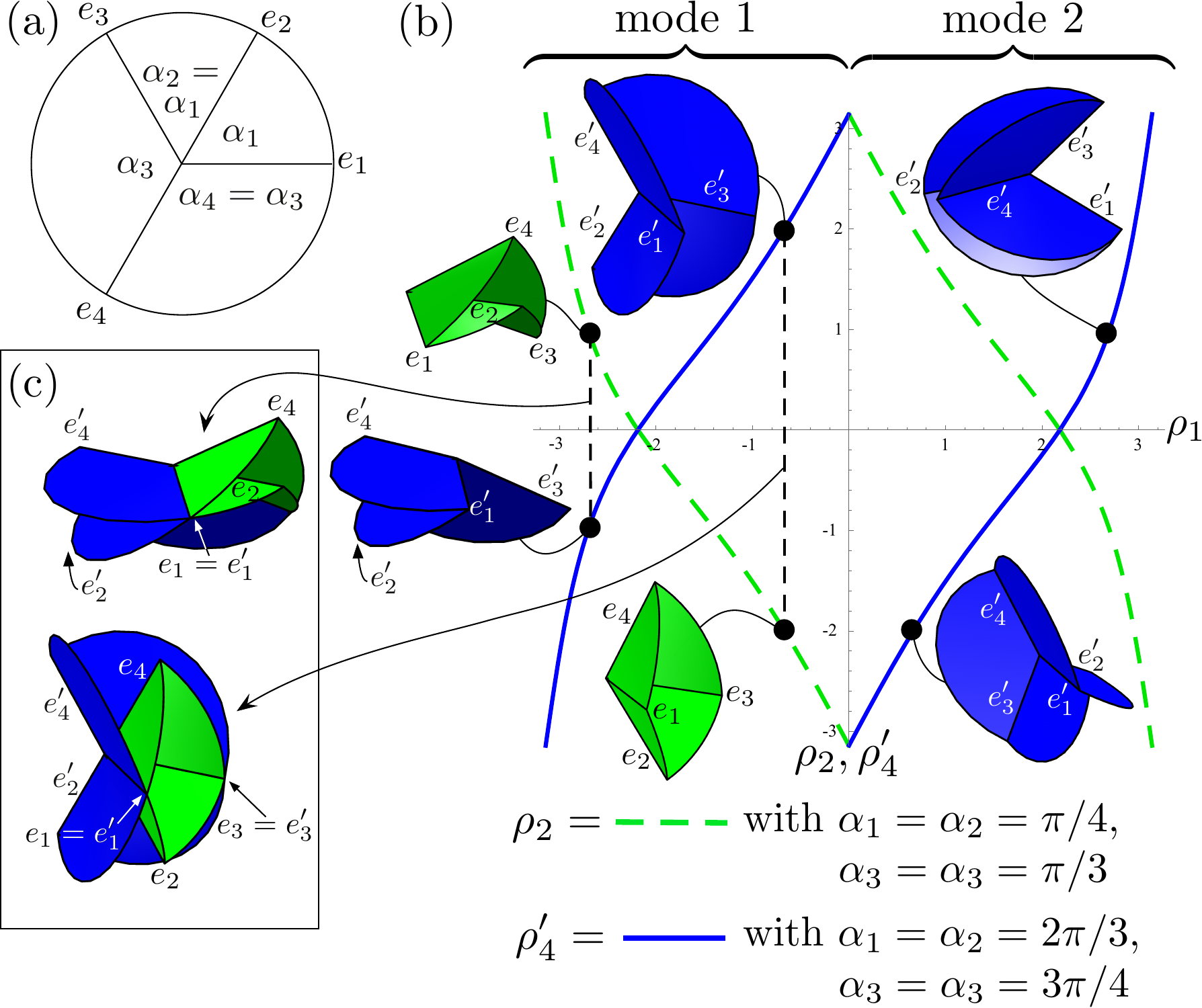}
    \caption{(a) A bird's foot vertex, where $\sum \alpha_i$ need not equal $2\pi$. (b) Graphs of the $(\rho_1, \rho_2)$ and $(\rho_1',\rho_4')$ relations from Equations~\eqref{ex2eq3} for a convex cone bird's foot with sector angles $(\pi/4,\pi/3)$ and a hyperbolic bird's foot with sectors $(2\pi/3,3\pi/4)$. (c) Placing the two vertices from (b) together to visualize the folding angle congruence.}
    \label{fig2.5}
\end{figure}

For another example showing the utility of the Theorem~\ref{thm1} equations, let us consider degree-4 vertices where two pairs of consecutive sector angles are equal, e.g., $\alpha_1=\alpha_2$ and $\alpha_3=\alpha_4$ (see Figure~\ref{fig2.5}(a)).  In the developable case, this is the flat-foldable ``bird's foot" vertex that forms the vertices in the much-studied Miura-ori crease pattern \cite{Silverberg1}. In this case previous results give us that the two folding modes in Theorem~\ref{thm1} become $\tan(\rho_2/2)=-\cos\alpha_1 \tan(\rho_1/2)$ for mode 1 and $\tan(\rho_1/2)=0$ for mode 2, implying that one of the folding modes has two creases (the left and right ``toes" of the bird's foot) being unfolded and the other two creases folding together in a straight line.

Our equations can replicate this and go further for the non-developable case. Substituting $\alpha_1=\alpha_2$ and $\alpha_3=\alpha_4=\pi-\alpha_1$ with $i=2$ into Equation~\eqref{confeq1} yields
$\tan^2(\rho_2/2) = \tan^2(\rho_4/2)$, implying that $\rho_2$ and $\rho_4$ are congruent up to sign. Using Equation~\eqref{confeq1} with $i=1$ also gives us $\rho_1=\rho_3$ up to sign. Then substituting this case into Equation~\eqref{confeq2} with $i=4$ produces
\begin{equation}\label{ex2eq1}
\tan\frac{\rho_1}{2}\left( (\cos\alpha_1 - \cos(3\alpha_1))\tan\frac{\rho_1}{2} - 4\sin^2\alpha_1
\tan\frac{\rho_4}{2}\right) = 0.
\end{equation}
Therefore either $\rho_1=0$, implying $\rho_3=0$ and $\rho_2=\rho_4$ and we are just folding the straight line made by creases $e_2$ and $e_4$, or the other factor in Equation~\eqref{ex2eq1} is zero, giving us
$\tan(\rho_4/2)= \cos\alpha_1 \tan(\rho_1/2)$.  Then using Equation~\eqref{confeq2} with $i=1$ gives us $\tan(\rho_2/2)=-\cos\alpha_1 \tan(\rho_1/2)$, which exactly supports the results from Theorem~\ref{thm1} but with slightly more generality; if $\alpha_1<\pi/2$ then $\rho_1$ will have the opposite sign as $\rho_2$ and the same sign as $\rho_4$, whereas if $\alpha_1>\pi/2$ then the reverse is true, $\rho_1$ will have the same sign as $\rho_4$ and be opposite from $\rho_2$.

In the non-developable case, substituting $\alpha_1=\alpha_2$ and $\alpha_3=\alpha_4$ with $i=1$ and then $i=2$ into Equation~\eqref{confeq1} simplifies to
\begin{equation}\label{ex2eq2}
\tan^2\frac{\rho_1}{2} = \tan^2\frac{\rho_3}{2}\quad \mbox{ and } \quad \cos\frac{\rho_2}{2} = \frac{\sin \alpha_3}{\sin\alpha_1}\cos\frac{\rho_4}{2},
\end{equation}
while Equation~\eqref{confeq2} with $i=1$ and $i=4$ gives us
\begin{equation}\label{ex2eq3}
\tan\frac{\rho_2}{2} = \frac{\sin\alpha_1 \cot\alpha_3 + \cos\alpha_1\cos\rho_1}{\sin\rho_1}
\end{equation}
and
\begin{equation}\label{ex2eq3.5}
\tan\frac{\rho_4}{2} = \frac{\sin\alpha_3 \cot\alpha_1 + \cos\alpha_3\cos\rho_1}{\sin\rho_1}.
\end{equation}
To our knowledge, these are new folding angle equations for non-developable bird's foot-type vertices.  

Note that previous studies on rigid foldings of bird's foot vertices show that if all four creases are to be rigidly folded (with no folding angles constantly zero), then the ``left and right toe" creases $e_1, e_3$ in Figure~\ref{fig2.5}(a) must have the same MV parity \cite{Origametry}.  Therefore the $(\rho_1, \rho_3$) relation in Equation~\eqref{ex2eq2} implies $\rho_1=\rho_3$. 

Furthermore, the symmetry evident in the Equations~\eqref{ex2eq3} and \eqref{ex2eq3.5} can be exploited. If we substitute $\alpha_1=\pi-\alpha_3$ and $\alpha_3=\pi-\alpha_1$ into  Equation~\eqref{ex2eq3.5} we get
\begin{equation}\label{ex2eq4}
\tan\frac{\rho_4}{2} = -\frac{\sin\alpha_1 \cot\alpha_3 + \cos\alpha_1\cos\rho_1}{\sin\rho_1},
\end{equation}
which is exactly the $(\rho_1,\rho_2)$ relation in Equation~\eqref{ex2eq3} but with a sign difference.  Therefore if we let $C$ be an elliptic bird's foot degree-4 vertex with sector angles $(\alpha_1,\alpha_3)$, say with $\alpha_1< \alpha_3$ and folding angles $\rho_i$, and let $C'$ be a hyperbolic degree-4 bird's foot vertex with sector angles $(\pi-\alpha_3, \pi-\alpha_1)$ and folding angles $\rho_i'$, we will have $\rho_2=-\rho_4'$ and, similarly, $\rho_2'=\rho_4$.  An example of this is illustrated in Figure\ref{fig2.5}(b), showing the graphs of our folding angle equations, identical up to sign.  Geometrically this can be verified by placing the convex cone $C$ and hyperbolic vertex $C'$ together with $e_1=e_1'$ and $e_3=e_3'$, so that the folded structure is really two intersecting planes of paper folding along the straight lines $e_2\leftrightarrow e_4$ and $e_2'\leftrightarrow e_4'$, as shown in Figure~\ref{fig2.5}(c). This also proves that $\rho_1=\rho_1'$ and $\rho_3=\rho_3'$.  The above verifies, and offers alternate proofs of, the compatible kinematics of certain eggbox and Miura-ori crease patterns shown in \cite{Pratapa2019} as well as the nested convex cone and hyperbolic vertices of the ``zippered" origami tubes of \cite{Tachi15}.

\bibliography{gendeg4.bib}

\end{document}